\documentclass[fleqn,usenatbib,letters]{mnras}

\usepackage{comment}
\usepackage[T1]{fontenc}

\DeclareRobustCommand{\VAN}[3]{#2}
\let\VANthebibliography\thebibliography
\def\thebibliography{\DeclareRobustCommand{\VAN}[3]{##3}\VANthebibliography}

\defcitealias{2021arXiv211110145T}{T21}
\usepackage{graphicx}	
\usepackage{amsmath}	
\usepackage[dvipsnames]{xcolor}
\usepackage{txfonts}

\usepackage[dvipsnames]{xcolor}
\newcommand{\rev}[1]{\textcolor{black}{#1}}

\newcommand{\tacc}{t_{\rm{acc},0}}

\newcommand{\tdisp}{t_{\rm{disp}}}
\newcommand{\MaccMd}{\dot{M}_{\rm{*}}-M_{D}}
\newcommand{\tlt}{t_{\rm{lt}}}
\newcommand{\revmnras}[1]{\textcolor{black}{#1}}
\newcommand{\revmnrasbis}[1]{\textcolor{black}{#1}}
\title{MHD disc winds can reproduce fast disc dispersal and the correlation between accretion rate and disc mass in Lupus}
\author[]{
B. Tabone,$^{1}$\thanks{E-mail: tabone@strw.leidenuniv.nl}
G. P. Rosotti,$^{1,2}$
G. Lodato,$^{3}$
P. J. Armitage,$^{4,5}$
A. J. Cridland,$^{6}$
E. F. van Dishoeck$^{1,6}$
\\
$^{1}$Leiden Observatory, Leiden University, PO Box 9513, 2300 RA Leiden, The Netherlands \\
$^{2}$School of Physics and Astronomy, University of Leicester, Leicester LE1 7RH, UK\\
$^{3}$Dipartimento di Fisica dell'Università degli Studi di Milano, Via Celoria 16, Milano I-20133, Italy \\
$^{4}$Department of Physics and Astronomy, Stony Brook University, Stony Brook, NY 11794, USA\\
$^{5}$Center for Computational Astrophysics, Flatiron Institute, New York, NY 10010, USA\\
$^{6}$Max-Planck-Institut für Extraterrestrische Physik, Giessenbachstrasse 1, D-85748 Garching bei München, Germany
}

\date{Accepted XXX. Received YYY; in original form ZZZ}

\pubyear{2021}

\begin{document}
\label{firstpage}
\pagerange{\pageref{firstpage}--\pageref{lastpage}}
\maketitle

\begin{abstract}
The final architecture of planetary systems depends on the extraction of angular momentum and mass-loss processes of the discs in which they form.
Theoretical studies proposed that magnetized winds launched from the discs (MHD disc winds) could govern accretion and disc dispersal. In this work, we revisit the observed disc demographics in the framework of MHD disc winds, combining analytical solutions of disc evolution and a disc population synthesis approach. We show that MHD disc winds alone can account for both disc dispersal and accretion properties. The decline of disc fraction over time is reproduced by assuming that the initial accretion timescale (a generalization of the viscous timescale) varies from disc to disc and that the decline of the magnetic field strength is slower than that of the gas. The correlation between  accretion rate and disc mass, and the dispersion of the data around the mean trend as observed in Lupus is then naturally reproduced. The model also accounts for the rapidity of the disc dispersal. This paves the way for planet formation models in the paradigm of wind-driven accretion.
\end{abstract}



\section{Introduction}
Unveiling the physical processes that regulate disc evolution is crucial to understand the emergence of the diversity and the habitability of exoplanets \citep{2016JGRE..121.1962M}. Extensive surveys from the UV to the millimeter have shown two essential features of disc evolution: (i) discs, as identified from their infrared (IR) excess, are accreting, implying a transport of angular momentum, and (ii) discs are dispersed after a few Myr in a short timescale ($\simeq 0.5~$Myr).

Over the past decades, these two observational facts have been explained by two distinct processes in the "paradigm of viscous discs" \citep{1973A&A....24..337S,1974MNRAS.168..603L}. On the one hand, the magneto-rotational instability, or perhaps other hydrodynamical instabilities would act as a viscosity, transporting the angular momentum in the radial direction \citep[][and references therein]{2011ARA&A..49..195A}. On the other hand, UV photons or X-rays would launch hydrodynamical winds, a.k.a photoevaporative winds, that would quickly disperse the disc \citep[][]{2014prpl.conf..475A}.

The new generation of telescopes revolutionised our view on disc evolution thanks to complete surveys of star-forming regions. ALMA has provided continuum fluxes, a proxy for disc masses \citep[e.g.,][]{2016ApJ...828...46A,2016ApJ...831..125P}, and VLT-XShooter has measured stellar properties and accretion rates \citep[e.g.,][]{2017A&A...600A..20A,2020A&A...639A..58M}. Combining theses surveys, \citet{2016A&A...591L...3M} and \citet{2017ApJ...847...31M} found a correlation between accretion rate and disc mass in Lupus and Chamaeleon with a nearly linear relationship. Whereas such a correlation is expected in viscous models \citep{1998ApJ...495..385H,2017MNRAS.468.1631R}, the large scatter in the relation is in tension with viscous evolution. \citet{2017MNRAS.472.4700L} explained this scatter by assuming long viscous timescales and a dispersion in disc ages but under-predict it in the older region of Upper Scorpius \citep{2020A&A...639A..58M}. \citet{2020MNRAS.498.2845S} invoked fast dust radial drift to increase the scatter but this scenario is in tension with disc sizes measured from dust emission \citep{2021MNRAS.507..818T}.

Today, the paradigm of viscous disc evolution is challenged. ALMA observations suggest that at least at large radius ($> 20~$au), turbulence levels are too low to sustain disc accretion \citep{2016ApJ...816...25P,2018ApJ...856..117F}. Recent numerical simulations further demonstrate that MRI turbulence is quenched in the low-ionization regions of discs called dead-zones, where most of the planets are forming \citep[$\gtrsim 1$~au ;][]{1996ApJ...457..355G,2011ApJ...736..144B}. Following the pioneering work by \citet{1982MNRAS.199..883B}, the emerging paradigm proposes that efficient disc accretion is driven by magnetized winds launched from the disc surface along magnetic field lines \citep[MHD disc winds,][]{1997A&A...319..340F,2013ApJ...769...76B}. If disc accretion is driven by an MHD disc wind, every step of planet formation would be impacted. 
Yet, the majority of planet formation models rely on viscous disc models as disc demographics have almost exclusively been analysed in this framework. 

In this Letter, we use a disc population synthesis approach based on a simple disc evolution model presented in a companion paper \citep[][\citetalias{2021arXiv211110145T} hereafter]{2021arXiv211110145T} to reconsider the observed disc demographics in the framework of MHD disc winds. The model and the observational dataset are presented in Sec. \ref{sec:method} and the comparison of the model with disc dispersal and accretion properties are detailed in Sec. \ref{sec:results}. Our findings are summarized and discussed in Sec. \ref{sec:discussion}.
\vspace{-0.4cm}

\section{Method}
\label{sec:method}
\subsection{Disc population synthesis}

The evolution of an ensemble of discs is computed from a disc evolution model presented in \citetalias{2021arXiv211110145T}. In short, the disc is treated in a 1D approach by vertically averaging disc quantities. An MHD disc wind transporting angular momentum and mass is launched from the full extent of the disc. The wind torque and the mass-loss rate are parameterized using a generalization of the $\alpha$ Shakura-Sunyaev parameter, denoted as $\alpha_{DW}$, and the magnetic lever arm parameter denoted as $\lambda$. \rev{The correspondence between $\alpha_{DW}$ and other parameterizations of the wind torque \citep[e.g.,][]{2016A&A...596A..74S} can be found in \citetalias{2021arXiv211110145T}.} In this work, we neglect the viscous torque, i.e., we test the hypothesis that MHD winds alone can account for disc evolution and dispersal. Further assuming that $\alpha_{DW}$ and $\lambda$ are constant across the disc, analytical solutions for the evolution of a disc with a surface density profile of
\begin{equation}
\Sigma(r,t) = \Sigma_c(t) \left(r/r_c \right)^{-1+\xi} e^{-r/r_{\rm{c}}}
\end{equation}
are found, where $\Sigma_c(t)$ is a characteristic surface density, \mbox{$\xi=1/[2(\lambda-1)]$} is the mass ejection index, and $r_c$ is the disc characteristic size. In the absence of viscosity, $r_c$ is constant in time as no angular momentum is radially transported. The disc extends down to an inner radius $r_{in}$. In order to describe discs that disperse at a finite time, we used the $\Sigma_{\text{c}}$-dependent wind torque solutions of \citetalias{2021arXiv211110145T}, for which $\alpha_{DW}$ increases (implicitly) with time as
\begin{equation}
\alpha_{DW}(t) = \alpha_{DW}(0) \left( \Sigma_c(t)/\Sigma_c(0) \right)^{-\omega},
\end{equation}
where $\omega$ is a phenomenological parameter between 0 and 1 that quantifies the (unknown) dissipation of the magnetic field. For $\omega = 1$, the magnetic field strength in the disc is constant.

\begin{figure}
	\includegraphics[width=\columnwidth]{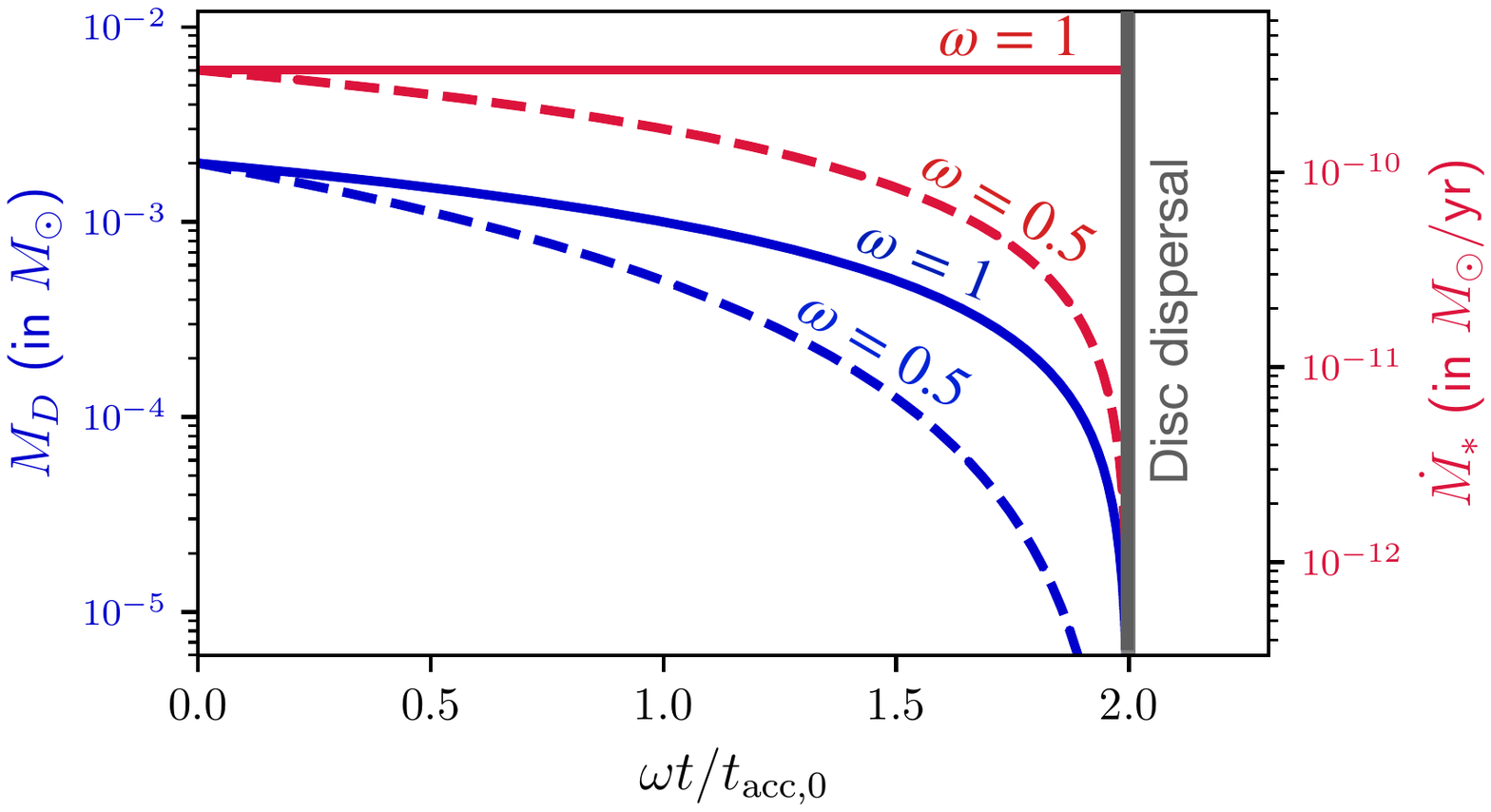}
    \caption{Evolution of the disc mass (in blue) and accretion rate (in red) for two values of the $\omega$ parameter. The accretion timescale is $\tacc=1$~Myr, the initial disc mass is $M_0=2 \times 10^{-3}M_{\odot}$, and $f_M=2$. For $\omega>0$ the disc is fully dispersed at a finite time $\tdisp$ (see Eq. \ref{eq:tdisp}). $\omega$ describes the time evolution of the magnetic field strength. High values of $\omega$ correspond to disc magnetic fields that decline more slowly, resulting in a shallower decline of the disc mass and accretion rate before dispersal.}
    \label{fig:intro-model}
\end{figure}

In this paper, we focus on the evolution of the disc mass $M_D(t)$ and the stellar accretion rate $\dot{M}_*(t)$ that read
\begin{equation}
\begin{split}
    M_{D}(t)& = M_0 \left( 1-\frac{\omega}{2 \tacc} t \right)^{1/\omega}, \\
    \dot{M}_{*}(t)& = \frac{M_0}{2\tacc (1+f_{M})} \left( 1-\frac{\omega}{2 \tacc}t \right)^{-1+1/\omega},
\end{split}
\label{eq:disc-mass}
\end{equation}
where 
\begin{equation}
    f_M  \equiv \dot{M}_W/\dot{M}_*= \left(r_c/r_{in} \right)^{\xi}-1
\end{equation}
is the mass ejection-to-accretion ratio, and
\begin{equation}
    \tacc \equiv \frac{r_c}{3 \epsilon_c c_{s,c} {\alpha_{DW}(0)}}
    \label{eq:tacc}
\end{equation}
is the initial accretion timescale which is a generalisation of the viscous timescale, where $\epsilon_c$ is the disc aspect ratio and $c_{s,c}$ is the sound speed at $r = r_c$. Therefore, $M_D(t)$ and $\dot{M}_*(t)$ are controlled by four independent parameters: $M_0$, $f_M$, $\tacc$, and $\omega$. The time evolution of the solutions is shown in Fig. \ref{fig:intro-model} and discussed below. \rev{Throughout this Letter, the initial time corresponds to the end of the Class I stage when the infall rate is much smaller than the disk accretion rate.}

From this simplified disc evolution model, we build a disc population model by randomly selecting initial disc mass $M_0$ and accretion timescale $\tacc$, and computing the evolution of the population. For sake of conciseness, for each synthetic population, $\omega$ and $f_M$ are fixed. The distribution of $\tacc$ is fitted from the observed disc fraction as detailed in Sec. \ref{subsec:discdisp}. The initial disc mass $M_0$ follows a log normal distribution with a dispersion of 1~dex. A dispersion in the predicted accretion rates of $0.45~$dex is added to account for short-term accretion variability as in \citet{2017MNRAS.472.4700L}.
\vspace{-0.4cm}
\subsection{Observational data}
\label{sec:observations}
The fractions of disc-bearing sources towards star-forming regions of different ages stem from the compilation by \citet{2010A&A...510A..72F} (see Fig. \ref{fig:fig-tacc-distrib}-a). New extensive ALMA and XShooter surveys open the possibility of accurately testing disc evolution models. In this work, disc masses and accretion rates in the Lupus star-forming region, where external photoevaporation is minimal, are collected from the compilation of \citet{2019A&A...631L...2M} who used distances from the Gaia DR2 \citep[][]{2018A&A...616A...1G}. The total disc masses (gas and dust) are estimated from dust continuum emission observed by ALMA \citep{2016ApJ...828...46A} by assuming a gas-to-dust ratio of 100, optically thin emission, a dust temperature of $20~$K, and a dust opacity of $2.3$ cm$^2$ g$^{-1}$ $(\nu/ 230~\rm{GHz})$. The dataset includes the young stellar objects (YSOs) in Lupus I-IV with stellar mass above $M_*>0.1~M_{\odot}$ with a completeness rate of $96 \%$. The accretion rates of the same sample are from a VLT-XShooter survey analysed by \citet{2014A&A...561A...2A,2017A&A...600A..20A}.

\vspace{-0.4cm}
\section{Results}
\label{sec:results}
\subsection{Disc dispersal}
\label{subsec:discdisp}

\begin{figure}
	\includegraphics[width=\columnwidth]{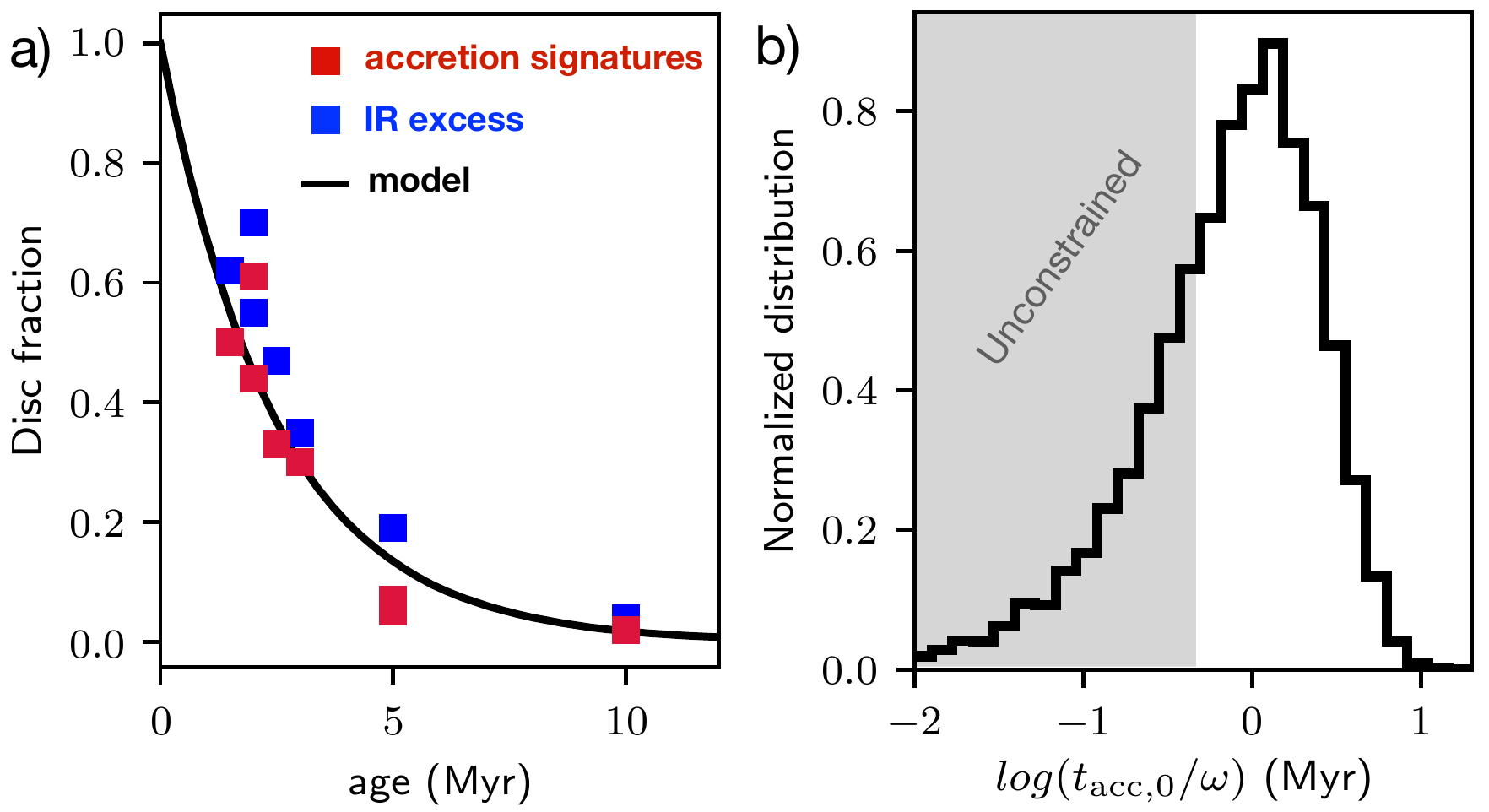}
    \caption{Distribution of initial accretion timescale $\tacc$ inferred from disc dispersal. \textit{a)} Fraction of sources with infrared excess and accretion signatures towards star-forming regions from \citet{2010A&A...510A..72F}, and the fit adopted in the present study. \textit{b)} Inferred distribution of $\tacc$. The x-axis is normalized by $\tacc/\omega$ such that the plotted distribution does not depends on $\omega$.}
    \label{fig:fig-tacc-distrib}
\end{figure}

Extensive surveys show that the fraction of sources with infrared excess or accretion signatures drops with the cluster age over a typical time of about $2-3~$Myr (see Fig. \ref{fig:fig-tacc-distrib}-a). Our disc evolution model predicts that a disc is fully dispersed after (see Fig. \ref{fig:intro-model}) 
\begin{equation}
    t_{\rm{disp}} = 2 \tacc/\omega.
    \label{eq:tdisp}
\end{equation}
We shall insist here that disc dispersal is not a result of the mass lost in the wind, but of the absence of disc spreading and of the increase in $\alpha_{DW}$ over time. Indeed, the disc dispersal time $t_{\rm{disp}}$ does not depend on the value of $f_{M}$, that is the fraction of mass ejected in the wind compared to that accreted onto the star. 
Following Eq. (\ref{eq:tdisp}), the decline of disc fraction with age can be interpreted as the result of a distribution of the initial accretion timescale $\tacc$. For example, the fact that $30 \%$ of YSOs bear a disc in the $3$ Myr old $\sigma$-Ori cluster points towards a population that had initially $30 \%$ of its discs born with $\tdisp > 3~$Myr. Following this approach, we assume that all the clusters studied by \citet{2010A&A...510A..72F} had initially the same distribution of $\tacc$ (or equivalently, $\tdisp$). We further assume that the transition between "disc-bearing" (Class II) and "disc-less" (Class III) stage occurs at $t=t_{\rm{disp}}$. The initial distribution of $\tacc$ required to fit the disc frequency with cluster age can then be derived from the fraction of disc-bearing sources denoted as $f_{D}$ (see supplementary material). One of the complications is that $f_{D}$ depends on the criterion used to measure the fraction of Class II sources. The dispersal time obtained from accretion signatures is shorter ($2.3~$Myr) than that of the inner disc traced by IR wavelengths ($3~$Myr). This effect might be the result of dust evolution and/or a difference between the sensitivity of the two diagnostics. We adopt here a characteristic disc dispersal time of $\tau=2.5$~Myr and a disc fraction of $f_D(t) =  e^{-t/\tau}$. 

The resulting distribution of $\tacc$ is shown in Fig. \ref{fig:fig-tacc-distrib}-b. The majority of discs are born with $\tacc$ of about $1.5/\omega$~Myr, which is, by construction, about half the disc dispersal time $\tau$. Because disc fraction is typically measured for clusters older than $\simeq 1~$Myrs, the distribution for short values of $\tacc$ ($\lesssim 0.5/\omega$~Myr) is poorly constrained. However, discs born with these short values of $\tacc$ are quickly dispersed and do not affect our predictions for ages above $\simeq 1$~Myr. In the following, all the synthetic populations follow the distribution of $\tacc$ shown in Fig. \ref{fig:fig-tacc-distrib}-b such that the fraction of disc-bearing sources in the synthetic populations always reproduces that observed. 

\subsection{Correlation between accretion rate and disc mass}
In this section, we show that given the distribution of $\tacc$ inferred from the disc fraction, the accretion properties observed in Lupus are naturally reproduced. We first adopt $\omega = 1$, which corresponds to discs of constant magnetic field strength. The only parameters that are left free are the distribution of initial disc masses $M_0$, and the value of $f_M$. \rev{In the model,} the evolution of disc mass depends only on $\tacc$ (see Eq. (\ref{eq:disc-mass}), and supplementary material). For $\omega=1$ the median mass of the synthetic population decreases by a factor of $3$ in 2 Myr. In order to reproduce the median disc mass inferred in Lupus, an initial median disc mass of $2 \times 10^{-3} M_{\odot}$ is then adopted.

\begin{figure}
	\includegraphics[width=\columnwidth]{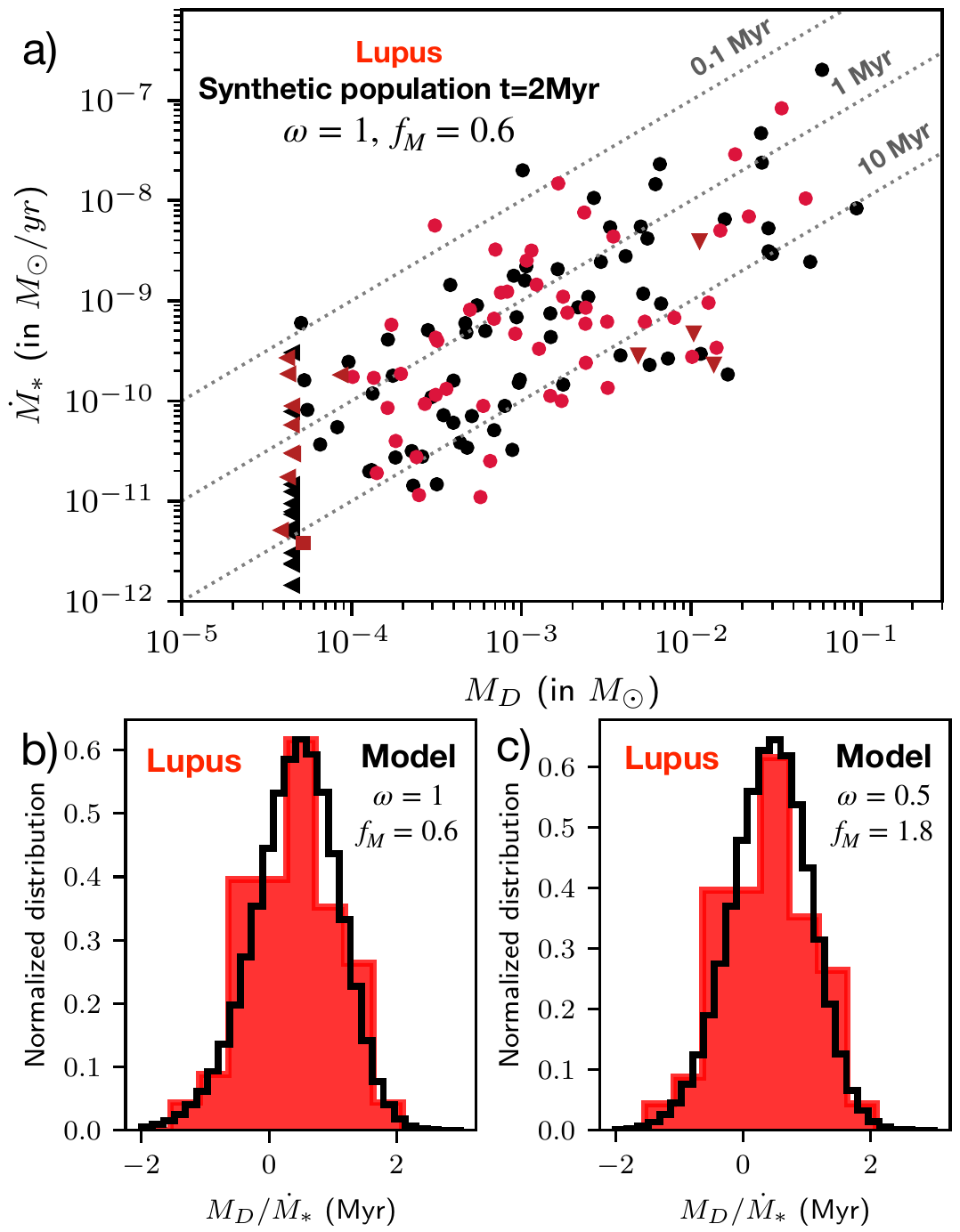}
    \caption{Correlation between accretion rate and disc mass. a) Location of the discs in the $\dot{M}_{*}-M_D$ plane after $2~$Myr for a synthetic population with $\omega=1$ and $f_M=0.6$ (black), and the Lupus disc population (red). b) The corresponding distribution of the disc lifetime $\tlt = M_D/\dot{M}_*$. In order to avoid statistical fluctuation, a sample of $10^6$ discs has been computed. Synthetic discs with a mass below the detection threshold of the ALMA survey are excluded. c) Same as panel b) but for $\omega = 0.5$ and $f_M=1.8$.}
    \label{fig:MD-Macc}
\end{figure}

We then follow the evolution of 200 discs, of which 90 have survived at the age of Lupus, in line with the disc fraction estimated in this cluster. Figure \ref{fig:MD-Macc}-a compares the accretion properties of a synthetic population after $2$~Myr for $f_M=0.6$ with the Lupus sample in the $\MaccMd$ plane. The model reproduces remarkably well the clustering of the data. We stress that the model has been constructed to reproduce disc fractions and the median disc mass. We have made no adjustment to reproduce the accretion rate distribution nor the dispersion in the data, so this is a significant achievement of the model. We recover the nearly linear relationship between accretion rates and disc masses found in the Lupus and Chamaeleon regions \citep{2016A&A...591L...3M,2017ApJ...847...31M}. Therefore, this correlation is not a distinctive feature of viscous evolution. In fact, the correlation found in our wind model is a consequence of the assumption that the distribution of $\tacc$ is independent of $M_0$. The latter assumption amounts to assume that disks of different masses are born with a similar distribution of magnetization and size. \revmnrasbis{It remains} to be determined by future work how much correlation between $\tacc$ and $M_0$ can be introduce to remain consistent with the data.

Even more striking is the agreement with the large dispersion of the data around the mean trend. As explained below, the predicted dispersion is the result of the dispersion in $\tacc$, which has been independently derived from the disc fraction. Because $M_D$ is on average proportional to $\dot{M}_*$, it is more convenient to investigate the accretion properties using the so-called observed disc lifetime:
\begin{equation}
    \tlt \equiv M_D/ \dot{M}_*.
    \label{eq:tlt-def}
\end{equation}
Indeed, in our model, $\tlt$ does not depend on the initial disc mass $M_0$ and the only free parameter that affects its value is $f_M$, $\tacc$ being set by disc fraction. The distribution of $\tlt$ shown in Fig. \ref{fig:MD-Macc}-b confirms the agreement between the model and the data. The median value of $\tlt$ is set by the distribution of $\tacc$ and the value of $f_M$. Increasing the ejection-to-accretion mass ratio $f_M$ lowers the accretion rates (Eq. (\ref{eq:disc-mass})), shifting the distribution of $\tlt$  to larger values by a factor $1+f_M$. In fact, the median disc lifetime of $\simeq 2.7~$Myr is recovered for $f_M=0.6$, a value in line with the mass-loss rates inferred from ALMA observations of MHD disc wind candidates \citep[e.g.,][]{2018A&A...618A.120L,2020A&A...634L..12D,2020A&A...640A..82T}.

The observed dispersion in $\tlt$ of $0.8$~dex around the median value is also reproduced, yet somewhat underestimated by the model ($0.65$ dex). Still, the predicted dispersion is a conservative value since any other processes affecting the estimates of the disc quantities are not included (apart from short-term variability of $\dot{M}_*$). The predicted dispersion in $\tlt$ is only set by the dispersion in $\tacc$, which stems from disc fraction. Therefore, our prediction is robust and suggests that the spread in the $\MaccMd$ plane, that reflects the extraction of angular momentum, is profoundly connected to disc dispersal time.

Interestingly, we find that the predicted distribution of $\tlt$ does not depend on time (see suppl. material). This result is in stark contrast with viscous evolution models that predict a decline of the dispersion and of the median disc lifetime with the age of the cluster \citep{1998ApJ...495..385H,2017MNRAS.472.4700L}. The dispersion observed in the older region Upper Sco ($\simeq 6~$Myr) could be reproduced by our simple model and dust evolution is not required to account for it, in contrast with viscous models \citep{2020MNRAS.498.2845S}. However, the comparison with Upper Sco is beyond the scope of this paper as dust evolution might still affect the gas-to-dust ratio and so the median value of $\tlt$.

We now explore a more generic case of $\omega < 1$ to demonstrate that there is no requirement in the model to fine tune $\omega$. For a smaller values of $\omega$, the disc mass and accretion rate of an individual disc experience a steeper drop before being dispersed (see Fig. \ref{fig:intro-model}). For $\omega=0.5$, we adopt a median disc mass of $4\times 10^{-3} M_{\odot}$ to reproduce than observed in Lupus. \rev{Interestingly, this value is more in line with the median mass of Class I discs than in the case $\omega=1$ \citep[when assuming a similar dust opacity coefficient as in Lupus, see][]{2020A&A...640A..19T,2020ApJ...890..130T}.}
As in the case $\omega=1$, the nearly linear relationship between $M_D$ and $\dot{M}_*$ is recovered. The median disc lifetime is well reproduced for $f_M = 1.8$ (Fig. \ref{fig:MD-Macc}-c), a value that is higher than in the case $\omega =1$. In fact, for lower values of $\omega$, the accretion timescale $\tacc$ required to fit disc dispersal is reduced by a factor $\omega$ (Fig. \ref{fig:fig-tacc-distrib}-b). Since $\dot{M}_* \propto \tacc^{-1}$ (Eq. \ref{eq:disc-mass}), lower values of $\omega$ lead to higher values of $\dot{M}_*$. In order to remain compatible with the observed median disc lifetime $M_D/\dot{M}_*$, a higher value of $f_M$ is required.

At this stage, any value of $\omega$ seems to reproduce the data. However, for $\omega < 0.5$, our criterion that the transition between Class II and Class III stage corresponds to $t=\tdisp$ fails. In fact, for these low values of $\omega$, disc masses and accretion rates smoothly drop below the detection limits of the surveys before $t=\tdisp$. In that case, our simplified approach provides a lower limit on $\tacc$. A more sophisticated analysis of detection thresholds, which is beyond the scope of the present paper, has to be carried out.
\vspace{-0.2cm}
\subsection{Rapidity of the disc dispersal}

\begin{figure}
	\includegraphics[width=0.91\columnwidth]{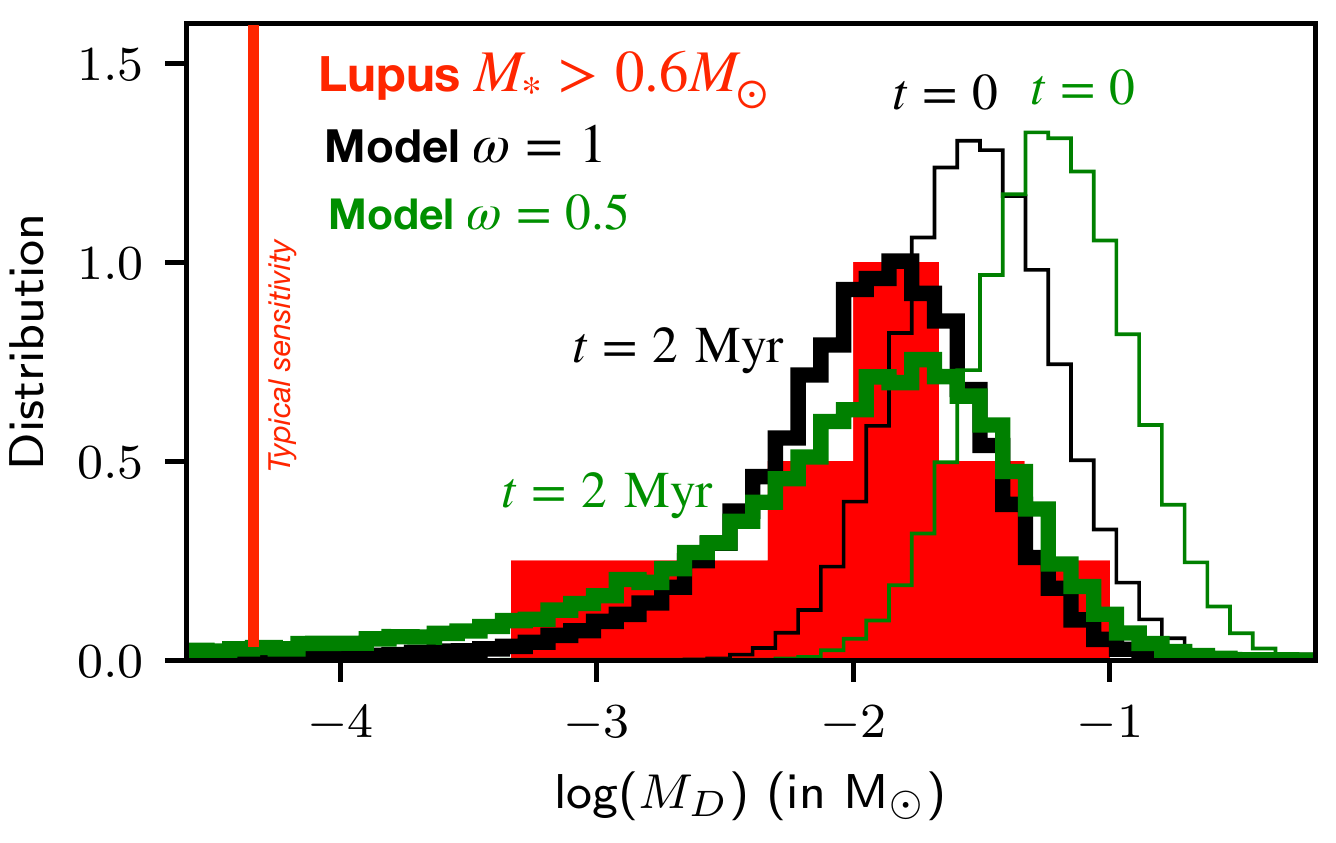}
	\vspace{-0.3em}
    \caption{Rapidity of disc dispersal as probed by the distribution of $M_{D}$ for stellar masses above $0.6 M_{\odot}$ inferred in Lupus (in red) compared to that predicted by our synthetic population for $\omega=1$ and $0.5$ (in black and green thick lines respectively). The initial distributions of disc mass for the two models are in thin lines. The scarcity of disc masses in the range $3\times 10^{-5} - 10^{-3}M_{\odot}$ demonstrates that our model reproduces the rapidity of the disc dispersal.}
    \label{fig:disc-disp}
\end{figure}

Disc dispersal is a fast process: after a few Myr, the disc is quickly dispersed within $\lesssim 0.5$ Myr. In this section, we show that without any further adjustment of $\tacc$ or $f_M$, the rapidity of disc dispersal is reproduced. To do so, we focus on the discs around the most massive stars, because only in this case is the observational mass sensitivity enough to detect all the Class II discs \citep{2021MNRAS.500.4878L}.

Figure \ref{fig:disc-disp} (red histogram) shows that discs observed in Lupus around stars more massive than $0.6 M_{\odot}$ are confined to high mass, typically above $M_D \simeq 10^{-3} M_{\odot}$, and far above the detection threshold of the survey. This can be interpreted as the footprint of a rapid dispersal of the disc: prior to its full dispersal, a disc runs along the x-axis of Fig. \ref{fig:disc-disp} as its mass drops. If discs were to disperse slowly, they would have been distributed all the way down to the detection threshold, which is not what is observed.

To compare our model to this sub-sample, we ran a synthetic population that describes discs around the most massive stars by increasing the initial median disc mass to $3 \times 10^{-2}/\omega~M_{\odot}$ to reproduce that observed around stars with $M_*\ge 0.6 M_{\odot}$. An initial dispersion of 0.3~dex is also adopted to reproduce the observed spread in $M_D$. This is lower than the dispersion adopted to reproduce the full Lupus sample. In fact, in the latter, the high dispersion in $M_D$ is driven by the distribution of stellar mass of the sample in virtue of the correlation between the stellar and disc mass \citep{2016ApJ...831..125P}. 

Figure \ref{fig:disc-disp} shows the predicted distribution of $M_D$ at $t=0$ and at the age of Lupus. For the case $\omega=1$ (black histograms), while the distribution of disc masses is initially symmetric (see thin lines), by the age of Lupus the distribution has acquired a tail at low disc masses, corresponding to discs that are being dispersed.
Overall, the distribution of disc mass in remarkably well reproduced. In particular, the scarcity of discs around $10^{-4} M_{\odot}$ is recovered. This demonstrates that the dispersal process predicted by our MHD wind model is fast enough. For lower values of $\omega$, the tail is somewhat more pronounced (see thick green histogram) as the mass of an individual disc declines more slowly prior to the full dispersal (see Fig. \ref{fig:intro-model}). Still the model with $\omega = 0.5$ reproduces well the data. For even lower values of $\omega$, the model is in tension with the data as the occurrence of low mass discs increases. However, as mentioned above, more sophisticated models have to be built to explore this parameter space.
\vspace{-0.7cm}

\section{Concluding remarks}
\label{sec:discussion}

In this work, we show that the essential features of disc evolution as probed by the observations are naturally reproduced by MHD wind-driven accretion. Starting with an initial distribution of accretion timescale $\tacc$, the frequency of disc-bearing sources with time is reproduced. Given this distribution and without further adjustment, the accretion properties of discs, as characterised by the correlation between accretion rate and disc mass, and the dispersion of the data around the mean trend, is naturally reproduced. A wind mass-loss rate of about the accretion rate ($f_M \simeq 1$) is required to reproduce the median $M_D/\dot{M}_*$ ratio. Finally, without any additional adjustment, the rapidity of the disc dispersal is accounted for. Therefore, the paradigmatic model of viscous evolution, for which disc accretion is governed by turbulence and dispersal by photoevaporative winds is not the only scenario that accounts for disc demographics, especially in regions without strong external photoevaporation such as Lupus. 

To first order, the distribution of $\tacc$ can be seen as the result of an underlying distribution of the disc size $r_c$ and the initial disc magnetization $\beta_0^{-1}$, that is the ratio between the thermal and the magnetic pressure in the mid-plane (see \citetalias{2021arXiv211110145T}, Eq. 70). Assuming a disc size of $r_c\simeq 50~$au, the distribution of $\tacc$ translates to a typical initial disc magnetization of $\beta_0 \simeq 10^{5}$ at $r=r_c$. Numerical simulations of disc formation \citep[e.g.,][]{2020A&A...635A..67H} running until the late Class I stage are needed to test this prediction and build a complete view of disc evolution, from their formation to their dispersal.

In this work, we estimated the disc gas mass from the (sub)millimeter dust emission assuming a standard gas-to-dust ratio of 100 and standard optical properties \citep{1990AJ.....99..924B}. 
Radial drift makes the dust-to-gas ratio decrease with time so that the gas mass may be underestimated. Modelling of dust coagulation and radial drift \citep{2020A&A...633A.114S} shows however that at the age of Lupus the dust masses likely remain better indicators of the total mass than alternative tracers such as CO isotopologues. If disc masses are higher than derived in this work, the estimated distribution of disc lifetimes $M_D/\dot{M}_*$ shown in Fig. \ref{fig:MD-Macc}-b-c would be shifted to higher values. Our model would still be able to reproduce the data with higher values of $f_M$, and higher initial median disc masses. We also note that if winds carry a significant fraction of mass ($f_M \gtrsim 1$), as suggested by this work, they would tend to increase the dust-to-gas ratio as winds are launched from the disc upper layers that are depleted in dust due to dust settling \rev{\citep{2016ApJ...821....3M,2019ApJ...882...33G}}. We plan to study these effects in future papers.

One of the most discriminant features between wind-driven and viscosity-driven accretion is the absence of disc spreading in the former scenario. ALMA surveys of the gas are required to evidence a possible disc viscous spreading \citep{2020A&A...640A...5T}. A deep systematic search for MHD disc winds launched from the bulk part of the discs has also to be conducted to determine if the MHD disc wind candidates unveiled in two YSOs, namely HH30 and HD162396 are common \citep[][Booth et al. 2021]{2018A&A...618A.120L}. Whereas  near-infrared and optical lines hint at the ubiquity of such winds \citep[e.g.,][]{2011ApJ...733...84P,2020A&A...643A..32G}, spatially and spectrally resolved observations, for example with ALMA, are required to infer the mass and angular momentum effectively extracted by the winds \citep{2020A&A...640A..82T}. Sources with the highest accretion rates, for which MHD disc winds are believed to be rich in CO should be prime targets \citep{2012A&A...538A...2P,2019ApJ...874...90W}.

All in all this first study provides us with observationally tested disc evolution models that are required to study dust evolution and planet formation in the emerging paradigm of MHD disc winds.

\vspace{-1.6em}

\section*{Acknowledgements}
The authors thank the anonymous referee for their constructive comments. B.T. acknowledges support from the research programme DAN II with project number $614.001.751$, which is (partly) financed by the Dutch Research Council (NWO). G.R. acknowledges support from the Netherlands Organisation for Scientific Research (NWO, program number $016$.Veni.$192.233$) and from an STFC Ernest Rutherford Fellowship (grant number ST/T$003855/1$). We are grateful to C. F. Manara for sharing the Lupus data.

\vspace{-0.4cm}

\section*{Data Availability}
\vspace{-0.2cm}
The data underlying this article will be shared on reasonable request to the corresponding author.

\vspace{-0.6cm}



\bibliographystyle{mnras}
\bibliography{export-bibtex} 

\begin{thebibliography}{}
\makeatletter
\relax
\def\mn@urlcharsother{\let\do\@makeother \do\$\do\&\do\#\do\^\do\_\do\%\do\~}
\def\mn@doi{\begingroup\mn@urlcharsother \@ifnextchar [ {\mn@doi@}
  {\mn@doi@[]}}
\def\mn@doi@[#1]#2{\def\@tempa{#1}\ifx\@tempa\@empty \href
  {http://dx.doi.org/#2} {doi:#2}\else \href {http://dx.doi.org/#2} {#1}\fi
  \endgroup}
\def\mn@eprint#1#2{\mn@eprint@#1:#2::\@nil}
\def\mn@eprint@arXiv#1{\href {http://arxiv.org/abs/#1} {{\tt arXiv:#1}}}
\def\mn@eprint@dblp#1{\href {http://dblp.uni-trier.de/rec/bibtex/#1.xml}
  {dblp:#1}}
\def\mn@eprint@#1:#2:#3:#4\@nil{\def\@tempa {#1}\def\@tempb {#2}\def\@tempc
  {#3}\ifx \@tempc \@empty \let \@tempc \@tempb \let \@tempb \@tempa \fi \ifx
  \@tempb \@empty \def\@tempb {arXiv}\fi \@ifundefined
  {mn@eprint@\@tempb}{\@tempb:\@tempc}{\expandafter \expandafter \csname
  mn@eprint@\@tempb\endcsname \expandafter{\@tempc}}}

\bibitem[\protect\citeauthoryear{{Alcal{\'a}} et~al.,}{{Alcal{\'a}}
  et~al.}{2014}]{2014A&A...561A...2A}
{Alcal{\'a}} J.~M.,  et~al., 2014, \mn@doi [\aap]
  {10.1051/0004-6361/201322254}, \href
  {https://ui.adsabs.harvard.edu/abs/2014A&A...561A...2A} {561, A2}

\bibitem[\protect\citeauthoryear{{Alcal{\'a}} et~al.,}{{Alcal{\'a}}
  et~al.}{2017}]{2017A&A...600A..20A}
{Alcal{\'a}} J.~M.,  et~al., 2017, \mn@doi [\aap]
  {10.1051/0004-6361/201629929}, \href
  {https://ui.adsabs.harvard.edu/abs/2017A&A...600A..20A} {600, A20}

\bibitem[\protect\citeauthoryear{{Alexander}, {Pascucci}, {Andrews}, {Armitage}
   \& {Cieza}}{{Alexander} et~al.}{2014}]{2014prpl.conf..475A}
{Alexander} R.,  {Pascucci} I.,  {Andrews} S.,  {Armitage} P.,   {Cieza} L.,
  2014, in PPVI.

\bibitem[\protect\citeauthoryear{{Ansdell} et~al.,}{{Ansdell}
  et~al.}{2016}]{2016ApJ...828...46A}
{Ansdell} M.,  et~al., 2016, \mn@doi [\apj] {10.3847/0004-637X/828/1/46}, \href
  {https://ui.adsabs.harvard.edu/abs/2016ApJ...828...46A} {828, 46}

\bibitem[\protect\citeauthoryear{{Armitage}}{{Armitage}}{2011}]{2011ARA&A..49..195A}
{Armitage} P.~J.,  2011, \mn@doi [\araa] {10.1146/annurev-astro-081710-102521},
  \href {https://ui.adsabs.harvard.edu/abs/2011ARA&A..49..195A} {49, 195}

\bibitem[\protect\citeauthoryear{{Bai} \& {Stone}}{{Bai} \&
  {Stone}}{2011}]{2011ApJ...736..144B}
{Bai} X.-N.,  {Stone} J.~M.,  2011, \mn@doi [\apj]
  {10.1088/0004-637X/736/2/144}, \href
  {https://ui.adsabs.harvard.edu/abs/2011ApJ...736..144B} {736, 144}

\bibitem[\protect\citeauthoryear{{Bai} \& {Stone}}{{Bai} \&
  {Stone}}{2013}]{2013ApJ...769...76B}
{Bai} X.-N.,  {Stone} J.~M.,  2013, \mn@doi [\apj]
  {10.1088/0004-637X/769/1/76}, \href
  {https://ui.adsabs.harvard.edu/abs/2013ApJ...769...76B} {769, 76}

\bibitem[\protect\citeauthoryear{{Beckwith}, {Sargent}, {Chini}  \&
  {Guesten}}{{Beckwith} et~al.}{1990}]{1990AJ.....99..924B}
{Beckwith} S. V.~W.,  {Sargent} A.~I.,  {Chini} R.~S.,   {Guesten} R.,  1990,
  \mn@doi [\aj] {10.1086/115385}, \href
  {https://ui.adsabs.harvard.edu/abs/1990AJ.....99..924B} {99, 924}

\bibitem[\protect\citeauthoryear{{Blandford} \& {Payne}}{{Blandford} \&
  {Payne}}{1982}]{1982MNRAS.199..883B}
{Blandford} R.~D.,  {Payne} D.~G.,  1982, \mn@doi [\mnras]
  {10.1093/mnras/199.4.883}, \href
  {https://ui.adsabs.harvard.edu/abs/1982MNRAS.199..883B} {199, 883}

\bibitem[\protect\citeauthoryear{{Fedele}, {van den Ancker}, {Henning},
  {Jayawardhana}  \& {Oliveira}}{{Fedele} et~al.}{2010}]{2010A&A...510A..72F}
{Fedele} D.,  {van den Ancker} M.~E.,  {Henning} T.,  {Jayawardhana} R.,
  {Oliveira} J.~M.,  2010, \mn@doi [\aap] {10.1051/0004-6361/200912810}, \href
  {https://ui.adsabs.harvard.edu/abs/2010A&A...510A..72F} {510, A72}

\bibitem[\protect\citeauthoryear{{Ferreira}}{{Ferreira}}{1997}]{1997A&A...319..340F}
{Ferreira} J.,  1997, \aap, \href
  {https://ui.adsabs.harvard.edu/abs/1997A&A...319..340F} {319, 340}

\bibitem[\protect\citeauthoryear{{Flaherty}, {Hughes}, {Teague}, {Simon},
  {Andrews}  \& {Wilner}}{{Flaherty} et~al.}{2018}]{2018ApJ...856..117F}
{Flaherty} K.~M.,  {Hughes} A.~M.,  {Teague} R.,  {Simon} J.~B.,  {Andrews}
  S.~M.,   {Wilner} D.~J.,  2018, \mn@doi [\apj] {10.3847/1538-4357/aab615},
  \href {https://ui.adsabs.harvard.edu/abs/2018ApJ...856..117F} {856, 117}

\bibitem[\protect\citeauthoryear{{Gaia Collaboration} et~al.,}{{Gaia
  Collaboration} et~al.}{2018}]{2018A&A...616A...1G}
{Gaia Collaboration} et~al., 2018, \mn@doi [\aap]
  {10.1051/0004-6361/201833051}, \href
  {https://ui.adsabs.harvard.edu/abs/2018A&A...616A...1G} {616, A1}

\bibitem[\protect\citeauthoryear{{Gammie}}{{Gammie}}{1996}]{1996ApJ...457..355G}
{Gammie} C.~F.,  1996, \mn@doi [\apj] {10.1086/176735}, \href
  {https://ui.adsabs.harvard.edu/abs/1996ApJ...457..355G} {457, 355}

\bibitem[\protect\citeauthoryear{{Gangi} et~al.,}{{Gangi}
  et~al.}{2020}]{2020A&A...643A..32G}
{Gangi} M.,  et~al., 2020, \mn@doi [\aap] {10.1051/0004-6361/202038534}, \href
  {https://ui.adsabs.harvard.edu/abs/2020A&A...643A..32G} {643, A32}

\bibitem[\protect\citeauthoryear{{Giacalone}, {Teitler}, {K{\"o}nigl}, {Krijt}
  \& {Ciesla}}{{Giacalone} et~al.}{2019}]{2019ApJ...882...33G}
{Giacalone} S.,  {Teitler} S.,  {K{\"o}nigl} A.,  {Krijt} S.,   {Ciesla} F.~J.,
   2019, \mn@doi [\apj] {10.3847/1538-4357/ab311a}, \href
  {https://ui.adsabs.harvard.edu/abs/2019ApJ...882...33G} {882, 33}

\bibitem[\protect\citeauthoryear{{Hartmann}, {Calvet}, {Gullbring}  \&
  {D'Alessio}}{{Hartmann} et~al.}{1998}]{1998ApJ...495..385H}
{Hartmann} L.,  {Calvet} N.,  {Gullbring} E.,   {D'Alessio} P.,  1998, \mn@doi
  [\apj] {10.1086/305277}, \href
  {https://ui.adsabs.harvard.edu/abs/1998ApJ...495..385H} {495, 385}

\bibitem[\protect\citeauthoryear{{Hennebelle}, {Commer{\c{c}}on}, {Lee}  \&
  {Charnoz}}{{Hennebelle} et~al.}{2020}]{2020A&A...635A..67H}
{Hennebelle} P.,  {Commer{\c{c}}on} B.,  {Lee} Y.-N.,   {Charnoz} S.,  2020,
  \mn@doi [\aap] {10.1051/0004-6361/201936714}, \href
  {https://ui.adsabs.harvard.edu/abs/2020A&A...635A..67H} {635, A67}

\bibitem[\protect\citeauthoryear{{Lodato}, {Scardoni}, {Manara}  \&
  {Testi}}{{Lodato} et~al.}{2017}]{2017MNRAS.472.4700L}
{Lodato} G.,  {Scardoni} C.~E.,  {Manara} C.~F.,   {Testi} L.,  2017, \mn@doi
  [\mnras] {10.1093/mnras/stx2273}, \href
  {https://ui.adsabs.harvard.edu/abs/2017MNRAS.472.4700L} {472, 4700}

\bibitem[\protect\citeauthoryear{{Louvet}, {Dougados}, {Cabrit}, {Mardones},
  {M{\'e}nard}, {Tabone}, {Pinte}  \& {Dent}}{{Louvet}
  et~al.}{2018}]{2018A&A...618A.120L}
{Louvet} F.,  {Dougados} C.,  {Cabrit} S.,  {Mardones} D.,  {M{\'e}nard} F.,
  {Tabone} B.,  {Pinte} C.,   {Dent} W.~R.~F.,  2018, \mn@doi [\aap]
  {10.1051/0004-6361/201731733}, \href
  {https://ui.adsabs.harvard.edu/abs/2018A&A...618A.120L} {618, A120}

\bibitem[\protect\citeauthoryear{{Lovell} et~al.,}{{Lovell}
  et~al.}{2021}]{2021MNRAS.500.4878L}
{Lovell} J.~B.,  et~al., 2021, \mn@doi [\mnras] {10.1093/mnras/staa3335}, \href
  {https://ui.adsabs.harvard.edu/abs/2021MNRAS.500.4878L} {500, 4878}

\bibitem[\protect\citeauthoryear{{Lynden-Bell} \& {Pringle}}{{Lynden-Bell} \&
  {Pringle}}{1974}]{1974MNRAS.168..603L}
{Lynden-Bell} D.,  {Pringle} J.~E.,  1974, \mn@doi [\mnras]
  {10.1093/mnras/168.3.603}, \href
  {https://ui.adsabs.harvard.edu/abs/1974MNRAS.168..603L} {168, 603}

\bibitem[\protect\citeauthoryear{{Manara} et~al.,}{{Manara}
  et~al.}{2016}]{2016A&A...591L...3M}
{Manara} C.~F.,  et~al., 2016, \mn@doi [\aap] {10.1051/0004-6361/201628549},
  \href {https://ui.adsabs.harvard.edu/abs/2016A&A...591L...3M} {591, L3}

\bibitem[\protect\citeauthoryear{{Manara}, {Mordasini}, {Testi}, {Williams},
  {Miotello}, {Lodato}  \& {Emsenhuber}}{{Manara}
  et~al.}{2019}]{2019A&A...631L...2M}
{Manara} C.~F.,  {Mordasini} C.,  {Testi} L.,  {Williams} J.~P.,  {Miotello}
  A.,  {Lodato} G.,   {Emsenhuber} A.,  2019, \mn@doi [\aap]
  {10.1051/0004-6361/201936488}, \href
  {https://ui.adsabs.harvard.edu/abs/2019A&A...631L...2M} {631, L2}

\bibitem[\protect\citeauthoryear{{Manara} et~al.,}{{Manara}
  et~al.}{2020}]{2020A&A...639A..58M}
{Manara} C.~F.,  et~al., 2020, \mn@doi [\aap] {10.1051/0004-6361/202037949},
  \href {https://ui.adsabs.harvard.edu/abs/2020A&A...639A..58M} {639, A58}

\bibitem[\protect\citeauthoryear{{Miyake}, {Suzuki}  \& {Inutsuka}}{{Miyake}
  et~al.}{2016}]{2016ApJ...821....3M}
{Miyake} T.,  {Suzuki} T.~K.,   {Inutsuka} S.-i.,  2016, \mn@doi [\apj]
  {10.3847/0004-637X/821/1/3}, \href
  {https://ui.adsabs.harvard.edu/abs/2016ApJ...821....3M} {821, 3}

\bibitem[\protect\citeauthoryear{{Morbidelli} \& {Raymond}}{{Morbidelli} \&
  {Raymond}}{2016}]{2016JGRE..121.1962M}
{Morbidelli} A.,  {Raymond} S.~N.,  2016, \mn@doi [Journal of Geophysical
  Research (Planets)] {10.1002/2016JE005088}, \href
  {https://ui.adsabs.harvard.edu/abs/2016JGRE..121.1962M} {121, 1962}

\bibitem[\protect\citeauthoryear{{Mulders}, {Pascucci}, {Manara}, {Testi},
  {Herczeg}, {Henning}, {Mohanty}  \& {Lodato}}{{Mulders}
  et~al.}{2017}]{2017ApJ...847...31M}
{Mulders} G.~D.,  {Pascucci} I.,  {Manara} C.~F.,  {Testi} L.,  {Herczeg}
  G.~J.,  {Henning} T.,  {Mohanty} S.,   {Lodato} G.,  2017, \mn@doi [\apj]
  {10.3847/1538-4357/aa8906}, \href
  {https://ui.adsabs.harvard.edu/abs/2017ApJ...847...31M} {847, 31}

\bibitem[\protect\citeauthoryear{{Panoglou}, {Cabrit}, {Pineau Des For{\^e}ts},
  {Garcia}, {Ferreira}  \& {Casse}}{{Panoglou}
  et~al.}{2012}]{2012A&A...538A...2P}
{Panoglou} D.,  {Cabrit} S.,  {Pineau Des For{\^e}ts} G.,  {Garcia} P.~J.~V.,
  {Ferreira} J.,   {Casse} F.,  2012, \mn@doi [\aap]
  {10.1051/0004-6361/200912861}, \href
  {https://ui.adsabs.harvard.edu/abs/2012A&A...538A...2P} {538, A2}

\bibitem[\protect\citeauthoryear{{Pascucci} et~al.,}{{Pascucci}
  et~al.}{2016}]{2016ApJ...831..125P}
{Pascucci} I.,  et~al., 2016, \mn@doi [\apj] {10.3847/0004-637X/831/2/125},
  \href {https://ui.adsabs.harvard.edu/abs/2016ApJ...831..125P} {831, 125}

\bibitem[\protect\citeauthoryear{{Pinte}, {Dent}, {M{\'e}nard}, {Hales},
  {Hill}, {Cortes}  \& {de Gregorio-Monsalvo}}{{Pinte}
  et~al.}{2016}]{2016ApJ...816...25P}
{Pinte} C.,  {Dent} W.~R.~F.,  {M{\'e}nard} F.,  {Hales} A.,  {Hill} T.,
  {Cortes} P.,   {de Gregorio-Monsalvo} I.,  2016, \mn@doi [\apj]
  {10.3847/0004-637X/816/1/25}, \href
  {https://ui.adsabs.harvard.edu/abs/2016ApJ...816...25P} {816, 25}

\bibitem[\protect\citeauthoryear{{Pontoppidan}, {Blake}  \&
  {Smette}}{{Pontoppidan} et~al.}{2011}]{2011ApJ...733...84P}
{Pontoppidan} K.~M.,  {Blake} G.~A.,   {Smette} A.,  2011, \mn@doi [\apj]
  {10.1088/0004-637X/733/2/84}, \href
  {https://ui.adsabs.harvard.edu/abs/2011ApJ...733...84P} {733, 84}

\bibitem[\protect\citeauthoryear{{Rosotti}, {Clarke}, {Manara}  \&
  {Facchini}}{{Rosotti} et~al.}{2017}]{2017MNRAS.468.1631R}
{Rosotti} G.~P.,  {Clarke} C.~J.,  {Manara} C.~F.,   {Facchini} S.,  2017,
  \mn@doi [\mnras] {10.1093/mnras/stx595}, \href
  {https://ui.adsabs.harvard.edu/abs/2017MNRAS.468.1631R} {468, 1631}

\bibitem[\protect\citeauthoryear{{Sanchis} et~al.,}{{Sanchis}
  et~al.}{2020}]{2020A&A...633A.114S}
{Sanchis} E.,  et~al., 2020, \mn@doi [\aap] {10.1051/0004-6361/201936913},
  \href {https://ui.adsabs.harvard.edu/abs/2020A&A...633A.114S} {633, A114}

\bibitem[\protect\citeauthoryear{{Sellek}, {Booth}  \& {Clarke}}{{Sellek}
  et~al.}{2020}]{2020MNRAS.498.2845S}
{Sellek} A.~D.,  {Booth} R.~A.,   {Clarke} C.~J.,  2020, \mn@doi [\mnras]
  {10.1093/mnras/staa2519}, \href
  {https://ui.adsabs.harvard.edu/abs/2020MNRAS.498.2845S} {498, 2845}

\bibitem[\protect\citeauthoryear{{Shakura} \& {Sunyaev}}{{Shakura} \&
  {Sunyaev}}{1973}]{1973A&A....24..337S}
{Shakura} N.~I.,  {Sunyaev} R.~A.,  1973, \aap, \href
  {https://ui.adsabs.harvard.edu/abs/1973A&A....24..337S} {500, 33}

\bibitem[\protect\citeauthoryear{{Suzuki}, {Ogihara}, {Morbidelli}, {Crida}  \&
  {Guillot}}{{Suzuki} et~al.}{2016}]{2016A&A...596A..74S}
{Suzuki} T.~K.,  {Ogihara} M.,  {Morbidelli} A.,  {Crida} A.,   {Guillot} T.,
  2016, \mn@doi [\aap] {10.1051/0004-6361/201628955}, \href
  {https://ui.adsabs.harvard.edu/abs/2016A&A...596A..74S} {596, A74}

\bibitem[\protect\citeauthoryear{{Tabone} et~al.,}{{Tabone}
  et~al.}{2020}]{2020A&A...640A..82T}
{Tabone} B.,  et~al., 2020, \mn@doi [\aap] {10.1051/0004-6361/201834377}, \href
  {https://ui.adsabs.harvard.edu/abs/2020A&A...640A..82T} {640, A82}

\bibitem[\protect\citeauthoryear{{Tabone}, {Rosotti}, {Cridland}, {Armitage}
  \& {Lodato}}{{Tabone} et~al.}{2021}]{2021arXiv211110145T}
{Tabone} B.,  {Rosotti} G.~P.,  {Cridland} A.~J.,  {Armitage} P.~J.,   {Lodato}
  G.,  2021, arXiv e-prints, \href
  {https://ui.adsabs.harvard.edu/abs/2021arXiv211110145T} {p. arXiv:2111.10145}

\bibitem[\protect\citeauthoryear{{Tobin} et~al.,}{{Tobin}
  et~al.}{2020}]{2020ApJ...890..130T}
{Tobin} J.~J.,  et~al., 2020, \mn@doi [\apj] {10.3847/1538-4357/ab6f64}, \href
  {https://ui.adsabs.harvard.edu/abs/2020ApJ...890..130T} {890, 130}

\bibitem[\protect\citeauthoryear{{Toci}, {Rosotti}, {Lodato}, {Testi}  \&
  {Trapman}}{{Toci} et~al.}{2021}]{2021MNRAS.507..818T}
{Toci} C.,  {Rosotti} G.,  {Lodato} G.,  {Testi} L.,   {Trapman} L.,  2021,
  \mn@doi [\mnras] {10.1093/mnras/stab2112}, \href
  {https://ui.adsabs.harvard.edu/abs/2021MNRAS.507..818T} {507, 818}

\bibitem[\protect\citeauthoryear{{Trapman}, {Rosotti}, {Bosman}, {Hogerheijde}
  \& {van Dishoeck}}{{Trapman} et~al.}{2020}]{2020A&A...640A...5T}
{Trapman} L.,  {Rosotti} G.,  {Bosman} A.~D.,  {Hogerheijde} M.~R.,   {van
  Dishoeck} E.~F.,  2020, \mn@doi [\aap] {10.1051/0004-6361/202037673}, \href
  {https://ui.adsabs.harvard.edu/abs/2020A&A...640A...5T} {640, A5}

\bibitem[\protect\citeauthoryear{{Tychoniec} et~al.,}{{Tychoniec}
  et~al.}{2020}]{2020A&A...640A..19T}
{Tychoniec} {\L}.,  et~al., 2020, \mn@doi [\aap] {10.1051/0004-6361/202037851},
  \href {https://ui.adsabs.harvard.edu/abs/2020A&A...640A..19T} {640, A19}

\bibitem[\protect\citeauthoryear{{Wang}, {Bai}  \& {Goodman}}{{Wang}
  et~al.}{2019}]{2019ApJ...874...90W}
{Wang} L.,  {Bai} X.-N.,   {Goodman} J.,  2019, \mn@doi [\apj]
  {10.3847/1538-4357/ab06fd}, \href
  {https://ui.adsabs.harvard.edu/abs/2019ApJ...874...90W} {874, 90}

\bibitem[\protect\citeauthoryear{{de Valon}, {Dougados}, {Cabrit}, {Louvet},
  {Zapata}  \& {Mardones}}{{de Valon} et~al.}{2020}]{2020A&A...634L..12D}
{de Valon} A.,  {Dougados} C.,  {Cabrit} S.,  {Louvet} F.,  {Zapata} L.~A.,
  {Mardones} D.,  2020, \mn@doi [\aap] {10.1051/0004-6361/201936950}, \href
  {https://ui.adsabs.harvard.edu/abs/2020A&A...634L..12D} {634, L12}

\makeatother
\end{thebibliography}
\vspace{-0.5cm}
\section*{Supporting information}
Supplementary material is available at MNRAS online.

\vspace{-0.4cm}


\appendix

\section*{Supplementary material}

\section{Inferring the distribution of $\tacc$}
\label{app:distrib-tacc}
Assuming that the transition between "disc-bearing" (Class II) and "disc-less" (Class III) stage occurs at $t_{\rm{disp}}=2\tacc/\omega$, the cumulative distribution of $\tacc$ required to fit the disc frequency is then
\begin{equation}
    P(\tacc < t) = f_D(2 t / \omega),
\end{equation}
where $P(\tacc < t)$ is the probability to have discs with $\tacc$ shorter than $t$, and
$f_{D}$ the fraction of disc-bearing sources as a function of the cluster age. In this work, we adopt $f_D=\exp(-t/\tau)$, where $\tau=2.5~$Myr. This leads to the probability distribution of
\begin{equation}
    \frac{d P}{d \log(\tacc)} = \frac{ 2 \tacc}{\omega \tau} e^{- 2 \tacc / \omega \tau}
        \label{eq:distrib-tacc}
\end{equation}
shown in Fig. \ref{fig:fig-tacc-distrib}-b.

\section{Median disc mass and accretion rate}
\label{app:median-MD-Macc}

\begin{figure}
	\includegraphics[width=\columnwidth]{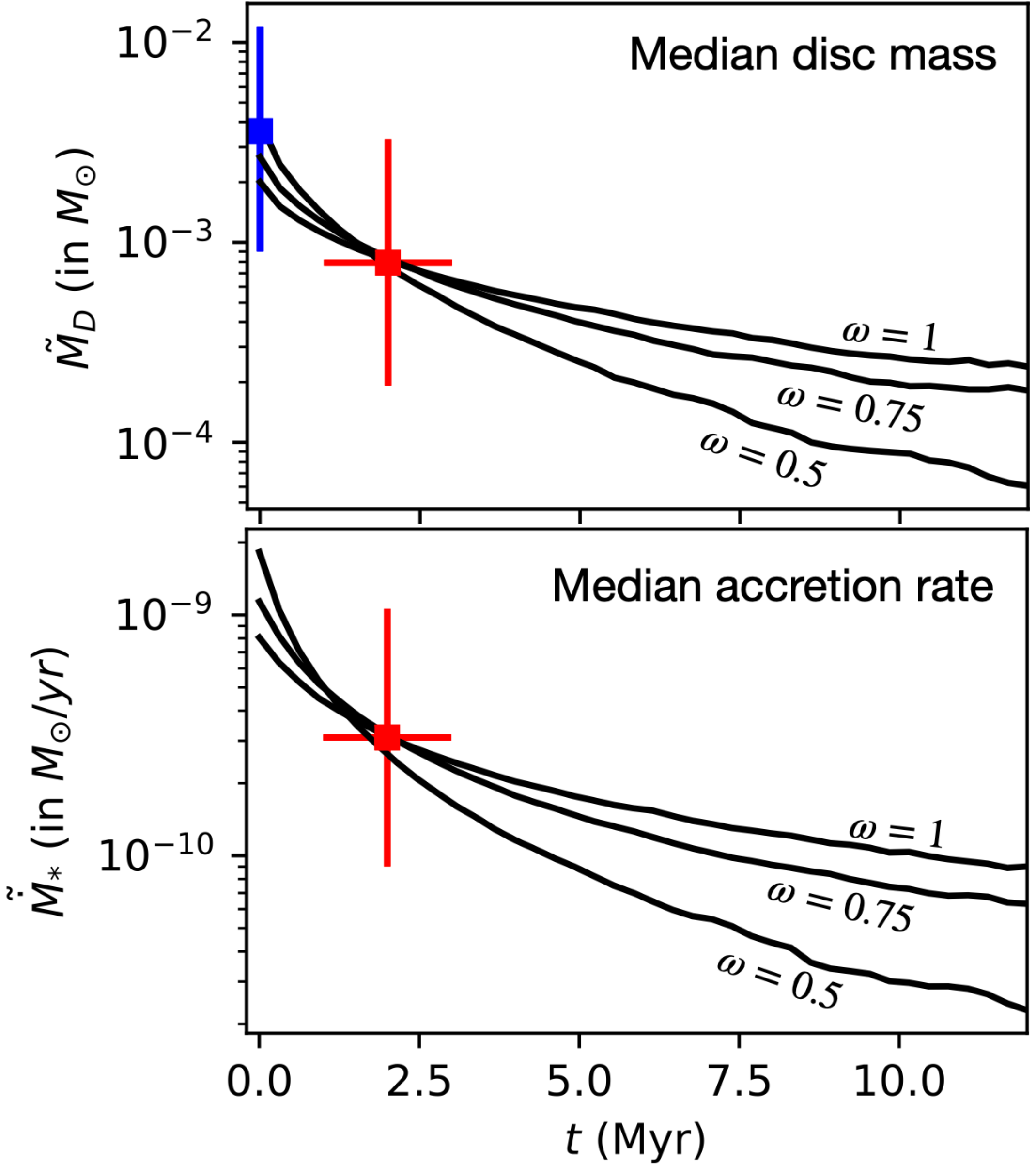}
    \caption{Evolution of the median disc mass and median accretion rate over the synthetic population of disc-bearing sources. The time evolution of these quantities is primarily set by the distribution of $\tacc$ \revmnrasbis{given in Eq. \ref{eq:distrib-tacc}}. The absolute value of $\tilde{M}_D$ and $\tilde{\dot{M}}_*$ can be rescaled by the initial median disc mass. $\tilde{\dot{M}}_*$ is also scaled by $1/(1+f_M)$. As an illustration, we adopt an initial median disc mass of $2\times 10^{-3}/\omega~M_{\odot}$ and $f_M=1.2/\omega-0.6$. \revmnrasbis{In order to reduce statistical noise, a sample of $10^6$ discs with a spread in initial disc mass of $0.3$ dex has been computed.} The red crosses indicate the median values, and the first and third quartiles of the Lupus sample. \revmnras{The blue marker indicates the median, and the first and third quartiles of the Class I disc masses inferred by \citet{2020A&A...640A..19T} in Perseus using ALMA continuum fluxes and the same dust opacity law as for the Lupus sample (see Sec \ref{sec:observations}).}}
    \label{fig:evol-median-MD-Macc}
\end{figure}

The time evolution of the median disc mass $\tilde{M}_D$ and the median accretion rate $\tilde{\dot{M}}_*$ of the synthetic populations is shown in Fig. \ref{fig:evol-median-MD-Macc}. Given the distribution of $\tacc$ required to match the disc frequency with cluster age, the evolution of $\tilde{M}_D$ and $\tilde{\dot{M}}_*$ depends only on $\omega$. We shall insist here on the fact that the evolution of $\tilde{M}_D$ and $\tilde{\dot{M}}_*$ is the result of both the time evolution of individual discs and the selection processes (i.e., disc dispersal). An extreme example is provided by the evolution of the median accretion rate. In the case of $\omega=1$, the accretion rate of an individual disc is constant over time (see Fig. \ref{fig:intro-model}) but the median accretion rate declines with time (see Fig. \ref{fig:evol-median-MD-Macc}) as discs with higher accretion rates are removed first.

The effect of the initial mass $M_0$ of an individual disc is simply to rescale $M_D$ and $\dot{M}_*$ by a constant factor, and the effect of $f_M$ is to rescale $\dot{M}_*$ by a factor of $1/(1+f_M)$ (see Eq. (\ref{eq:disc-mass})). As the distribution of $M_0$ and $f_M$ are not correlated with $\tacc$, $\tilde{M}_D$ and $\tilde{\dot{M}}_*$ are simply rescaled accordingly. Figure \ref{fig:evol-median-MD-Macc} shows that the evolution of the median disc mass declines faster for lower values of $\omega$. Indeed, for lower values of $\omega$, the decline of the disc mass and accretion rate of an individual disc prior to dispersal is steeper (see Fig. \ref{fig:intro-model}).

\section{Distribution of the disc lifetime over time}
\label{app:disc-liftime}
We find that when the detection thresholds are not considered, the distribution of the disc lifetime as defined by $\tlt\equiv M_D/\dot{M}_*$ does not depend on time. This results might be surprising as the lifetime of an individual disc decreases linearly with time as (from Eqs. (\ref{eq:disc-mass}) and (\ref{eq:tlt-def})):
\begin{equation}
\tlt(t) = (1+f_M) \left( 2 \tacc - \omega t \right).
\label{eq:disc-lifetime}
\end{equation}
This apparent contradiction is due to a selection process. As the disc population evolves, the disc lifetime $\tlt$ of individual discs decreases. However, discs with smaller $\tacc$ so smaller $\tlt$ are removed first. In short, the evolution of individual discs tends to decrease the disc lifetime but disc dispersal tends to increase the average disc lifetime, resulting in an effectively constant distribution of $\tlt$. Quantitatively, discs born with a disc lifetime of $\tlt$ at a time $t$ were born with $\tacc = \frac{\tlt}{2 (1+f_M)} + \frac{\omega}{2} t$. Denoting $dP$ the probability to find discs with a disc lifetime between $\tlt$ and $\tlt+d\tlt$ at a time $t$, the distribution of disc lifetime is
\begin{equation}
\begin{split}
\frac{dP}{d \tlt} = &\frac{dP}{d\tacc} \frac{d\tacc}{d \tlt} \\
 = &\frac{1}{\omega \tau (1+f_M)} exp\left(-\frac{\tlt}{(1+f_M) \omega \tau}\right) f_D(t).
 \label{eq:appendix-distrib-tlt}
\end{split}
\end{equation}
Therefore, the distribution of $\tlt$ does not depend on time as the last term is the disc fraction that is simply a normalization factor. It also demonstrates the profound link between disc dispersal and disc accretion as the distribution of $\tlt$ is controlled by the dispersal time $\tau$. \rev{In particular, the median accretion timescale is proportional to $\tau \omega (1+f_M)$. However, we note that detection limits bias the observed distribution of $\tlt$ and introduce a time dependency. Still, we find that for the adopted initial distribution of disc mass and for $\omega \gtrsim 0.3$, the distribution does not depend significantly on time as detection rates are high in Lupus. We also recall that an additional dispersion of $0.45$~dex is added in our prediction to account for the effect of short-term variability of the stellar accretion rates.}

\bsp	
\label{lastpage}
\end{document}